\documentclass[12pt]{article}

\usepackage[a4paper,margin=1in]{geometry}
\usepackage{amsmath,amssymb,amsfonts,bm}
\usepackage{cite}
\usepackage{hyperref}
\usepackage{authblk}

\hypersetup{
    colorlinks=true,
    linkcolor=blue,
    citecolor=blue,
    urlcolor=blue
}

\newcommand{\Cres}{C_{\rm res}}
\newcommand{\TRVB}{T_{\rm RVB}}
\newcommand{\Tzero}{T_{0}}
\newcommand{\rhp}{r_{+}}
\newcommand{\XiQ}{\Xi}
\newcommand{\Area}{A_{+}}
\newcommand{\dd}{\mathrm{d}}
\newcommand{\ii}{\mathrm{i}}

\title{Dirac-Field Black Hole Entropy in \(f(Q)\) Gravity from the RVB Residue Method}

\author{
    Wen-Xiang Chen\\
    School of Electronic Information\\
    Guangzhou City University of Technology\\
    Department of Astronomy\\
    School of Physics and Materials Science\\
    Guangzhou University\\
    \texttt{wxchen4277@qq.com}
}

\date{\today}

\begin{document}

\maketitle

\begin{abstract}
We calculate the entropy of a Dirac quantum field near a static spherically symmetric black hole in \(f(Q)\) gravity by combining two ingredients. The first ingredient is the residue-based Robson--Villari--Biancalana method, in which the Hawking temperature is written as a surface-gravity term plus a residue-induced correction. The second ingredient is the thin-film or brick-wall state-counting method for a Dirac field in a near-horizon background. Starting from the \(f(Q)\)-deformed metric
\begin{equation}
    \dd s^{2}=-g(r)\dd t^{2}+\frac{\dd r^{2}}{g(r)}+r^{2}\dd\Omega^{2},
\end{equation}
we derive the Hamilton--Jacobi equation for the Dirac field, obtain the radial momentum, count the fermionic modes, and compute the free energy and entropy at the residue-corrected temperature. The final result shows that the Dirac-field entropy is still proportional to the horizon area after a proper cutoff is introduced, while the proportionality factor is corrected by the RVB residue through a cubic temperature factor. For the quadratic model \(f(Q)=Q+\alpha Q^{2}\), an explicit closed expression is obtained.
\end{abstract}


\section{Introduction}

Black hole entropy is one of the most direct bridges among gravity, quantum theory, and thermodynamics. The standard area law was proposed by Bekenstein and Hawking and was later related to Euclidean methods, Noether charge, and horizon thermodynamics \cite{Bekenstein1973,Hawking1975,Bardeen1973,GibbonsHawking1977,Wald1993,IyerWald1994}. Quantum matter fields near the horizon provide another statistical route to black hole entropy through the brick-wall or thin-film method \cite{tHooft1985,FrolovNovikov1993,SusskindUglum1994,Solodukhin2011}. In this picture, the entropy comes from the large density of quantum states close to the horizon, and a near-horizon cutoff is introduced to regulate the ultraviolet divergence.

The Dirac-field version of this calculation is especially useful because the leading WKB equation of a spinor field reduces to a Hamilton--Jacobi equation. Therefore, at leading order, the radial momentum of the Dirac field is controlled by the same near-horizon pole structure as scalar fields, while the fermionic statistics and spin degeneracy change the numerical coefficient \cite{Chandrasekhar1983,BirrellDavies1982,ParkerToms2009,Zhang2003Dirac}. This is the reason why the thin-film calculation of a Dirac field still leads to an entropy proportional to the horizon area after a suitable cutoff is chosen.

In the modified-gravity sector, \(f(Q)\) gravity is a nonmetricity-based extension of symmetric teleparallel gravity \cite{Nester1999,Jimenez2018Coincident,BeltranJimenez2019Trinity,Heisenberg2024Review}. Its gravitational dynamics are governed by a function of the nonmetricity scalar \(Q\). Black hole thermodynamics in such nonmetricity-based theories is subtle, because entropy may be interpreted either through an effective first-law reconstruction or through a Noether-charge-like analysis \cite{Heisenberg2022Wald,Hammad2019Teleparallel,Gomes2023EnergyEntropy}. 

The RVB method proposed by Robson, Villari, and Biancalana gives a complex-analytic way of extracting the Hawking temperature from the residue of a horizon pole \cite{Robson2019RVB}. In recent applications to \(f(Q)\) black holes, the RVB temperature can be written as the usual surface-gravity temperature plus a residue-induced shift \cite{Chen2025CJP,Chen2026Entropy}. The purpose of this paper is to combine that RVB-corrected temperature with the Dirac thin-film entropy calculation. The result should be interpreted as a test-field, near-horizon, thermodynamic entropy branch rather than a universal Noether-charge theorem.

\section{\(f(Q)\) black hole background and RVB temperature}

We use natural units
\begin{equation}
    G=\hbar=c=k_{\rm B}=1.
\end{equation}
The gravitational action of \(f(Q)\) gravity coupled to a Dirac field is written as
\begin{equation}
    I=\frac{1}{16\pi}\int \dd^{4}x\sqrt{-g}\,f(Q)+I_{\rm D},
\end{equation}
where \(Q\) is the nonmetricity scalar and \(I_{\rm D}\) is the Dirac-field action. In the static spherically symmetric sector, we take
\begin{equation}
    \dd s^{2}=-g(r)\dd t^{2}+\frac{\dd r^{2}}{g(r)}+r^{2}\left(\dd\theta^{2}+\sin^{2}\theta\,\dd\phi^{2}\right).
\end{equation}
The event horizon radius \(\rhp\) is determined by
\begin{equation}
    g(\rhp)=0.
\end{equation}

Following the residue-corrected \(f(Q)\) setup, we parameterize the metric function as
\begin{equation}
    g(r)=1-\frac{2M}{r}+\psi_{Q}(r),
\end{equation}
where \(\psi_{Q}(r)\) encodes the \(f(Q)\)-induced deformation. The horizon equation gives
\begin{equation}
    1-\frac{2M}{\rhp}+\psi_{Q}(\rhp)=0.
\end{equation}
Solving this equation for \(M\), we obtain
\begin{equation}
    M(\rhp)=\frac{\rhp}{2}\left[1+\psi_{Q}(\rhp)\right].
\end{equation}
Differentiating \(M(\rhp)\) with respect to \(\rhp\), we obtain
\begin{equation}
    \frac{\dd M}{\dd \rhp}
    =
    \frac{1}{2}\left[1+\psi_{Q}(\rhp)+\rhp\psi_{Q}'(\rhp)\right].
\end{equation}
For later convenience, define
\begin{equation}
    \XiQ(\rhp)
    =
    1+\psi_{Q}(\rhp)+\rhp\psi_{Q}'(\rhp).
\end{equation}
Then
\begin{equation}
    \frac{\dd M}{\dd \rhp}
    =
    \frac{1}{2}\XiQ(\rhp).
\end{equation}
The radial derivative of the metric function at the horizon is
\begin{equation}
    g'(\rhp)
    =
    \frac{1+\psi_{Q}(\rhp)}{\rhp}+\psi_{Q}'(\rhp).
\end{equation}
Using the definition of \(\XiQ(\rhp)\), this can be rewritten as
\begin{equation}
    g'(\rhp)
    =
    \frac{\XiQ(\rhp)}{\rhp}.
\end{equation}
Therefore the ordinary surface-gravity temperature is
\begin{equation}
    \Tzero(\rhp)
    =
    \frac{g'(\rhp)}{4\pi}
    =
    \frac{\XiQ(\rhp)}{4\pi \rhp}.
\end{equation}

In the RVB method, a simple horizon zero of \(g(r)\) becomes a simple pole of \(1/g(z)\) after complexification \(r\rightarrow z\). Hence
\begin{equation}
    \operatorname*{Res}_{z=\rhp}\frac{1}{g(z)}
    =
    \frac{1}{g'(\rhp)}.
\end{equation}
The corresponding inverse Hawking temperature can be written as
\begin{equation}
    \beta_{0}
    =
    4\pi \operatorname*{Res}_{z=\rhp}\frac{1}{g(z)}
    =
    \frac{4\pi}{g'(\rhp)}.
\end{equation}
Thus
\begin{equation}
    \Tzero
    =
    \beta_{0}^{-1}
    =
    \frac{g'(\rhp)}{4\pi}.
\end{equation}

For a more general residue correction, define a complex function \(F(z)\) associated with the metric function, the nonmetricity scalar, or another horizon-data function. The winding number is
\begin{equation}
    N_{\Gamma}
    =
    \frac{1}{2\pi \ii}\oint_{\Gamma}\frac{F'(z)}{F(z)}\,\dd z
    =
    \sum_{z_{k}\in \Gamma}
    \operatorname*{Res}_{z=z_{k}}\frac{F'(z)}{F(z)}.
\end{equation}
The residue-induced temperature shift is written as
\begin{equation}
    \Cres
    =
    \lambda_{\rm R} N_{\Gamma},
\end{equation}
where \(\lambda_{\rm R}\) is a normalization constant carrying the dimension of temperature. The RVB-corrected temperature is then
\begin{equation}
    \TRVB(\rhp)
    =
    \frac{g'(\rhp)}{4\pi}+\Cres.
\end{equation}
Using \(g'(\rhp)=\XiQ(\rhp)/\rhp\), this becomes
\begin{equation}
    \TRVB(\rhp)
    =
    \frac{\XiQ(\rhp)}{4\pi \rhp}+\Cres.
\end{equation}

\section{Geometric entropy reconstructed from the first law}

Before computing the Dirac-field entropy, it is useful to recall the geometric entropy branch generated by the same RVB temperature. The first law is assumed to be
\begin{equation}
    \dd M=\TRVB\,\dd S_{\rm geo}.
\end{equation}
Therefore
\begin{equation}
    \frac{\dd S_{\rm geo}}{\dd \rhp}
    =
    \frac{1}{\TRVB(\rhp)}
    \frac{\dd M}{\dd \rhp}.
\end{equation}
Substituting
\begin{equation}
    \frac{\dd M}{\dd \rhp}
    =
    \frac{1}{2}\XiQ(\rhp)
\end{equation}
and
\begin{equation}
    \TRVB(\rhp)
    =
    \frac{\XiQ(\rhp)}{4\pi \rhp}+\Cres,
\end{equation}
we obtain
\begin{equation}
    \frac{\dd S_{\rm geo}}{\dd \rhp}
    =
    \frac{\frac{1}{2}\XiQ(\rhp)}
    {\frac{\XiQ(\rhp)}{4\pi \rhp}+\Cres}.
\end{equation}
Multiplying numerator and denominator by \(4\pi \rhp\), one obtains
\begin{equation}
    \frac{\dd S_{\rm geo}}{\dd \rhp}
    =
    \frac{2\pi \rhp\,\XiQ(\rhp)}
    {\XiQ(\rhp)+4\pi \Cres \rhp}.
\end{equation}
Hence the residue-corrected geometric entropy is
\begin{equation}
    S_{\rm geo}(\rhp)
    =
    \int^{\rhp}
    \frac{2\pi u\,\XiQ(u)}
    {\XiQ(u)+4\pi \Cres u}\,\dd u
    +S_{0}.
\end{equation}
When \(\Cres=0\), the integrand reduces to \(2\pi u\), so
\begin{equation}
    S_{\rm geo}(\rhp)
    =
    \pi \rhp^{2}+S_{0}.
\end{equation}
Choosing \(S_{0}=0\), we recover the Bekenstein--Hawking area law
\begin{equation}
    S_{\rm BH}
    =
    \frac{\Area}{4},
    \qquad
    \Area=4\pi \rhp^{2}.
\end{equation}

For a small residue shift satisfying
\begin{equation}
    \left|4\pi \Cres \rhp\right|\ll \left|\XiQ(\rhp)\right|,
\end{equation}
we expand the denominator as
\begin{equation}
    \frac{1}{\XiQ(\rhp)+4\pi \Cres\rhp}
    =
    \frac{1}{\XiQ(\rhp)}
    \left[
    1-\frac{4\pi \Cres\rhp}{\XiQ(\rhp)}
    \right]
    +O(\Cres^{2}).
\end{equation}
Therefore
\begin{equation}
    \frac{\dd S_{\rm geo}}{\dd \rhp}
    =
    2\pi \rhp
    \left[
    1-\frac{4\pi \Cres\rhp}{\XiQ(\rhp)}
    \right]
    +O(\Cres^{2}).
\end{equation}
After integration, the first-order entropy is
\begin{equation}
    S_{\rm geo}(\rhp)
    =
    \frac{\Area}{4}
    -
    8\pi^{2}\Cres
    \int^{\rhp}
    \frac{u^{2}}{\XiQ(u)}\,\dd u
    +O(\Cres^{2}).
\end{equation}

\section{Dirac field in the \(f(Q)\) black hole background}

The Dirac action in curved spacetime is
\begin{equation}
    I_{\rm D}
    =
    \int \dd^{4}x\sqrt{-g}\,
    \bar{\Psi}
    \left[
    \ii\gamma^{a}e_{a}^{\ \mu}D_{\mu}-\mu
    \right]\Psi,
\end{equation}
where \(\mu\) is the Dirac mass, \(e_{a}^{\ \mu}\) is the tetrad, and \(D_{\mu}\) is the spinor covariant derivative. A convenient tetrad for the metric is
\begin{equation}
    e_{a}^{\ \mu}
    =
    \operatorname{diag}
    \left(
    \frac{1}{\sqrt{g(r)}},
    \sqrt{g(r)},
    \frac{1}{r},
    \frac{1}{r\sin\theta}
    \right).
\end{equation}
The Dirac equation is
\begin{equation}
    \left[
    \ii\gamma^{a}e_{a}^{\ \mu}D_{\mu}-\mu
    \right]\Psi=0.
\end{equation}

To obtain the leading WKB equation, take the ansatz
\begin{equation}
    \Psi
    =
    a(x)\exp\left(\frac{\ii}{\hbar}I(x)\right),
\end{equation}
where \(a(x)\) is a slowly varying spinor amplitude. Keeping only the leading order in \(\hbar\), the spin connection does not contribute to the principal Hamilton--Jacobi equation. Thus
\begin{equation}
    \left[
    \gamma^{a}e_{a}^{\ \mu}\partial_{\mu}I-\mu
    \right]a(x)=0.
\end{equation}
Multiplying by
\begin{equation}
    \left[
    \gamma^{b}e_{b}^{\ \nu}\partial_{\nu}I+\mu
    \right],
\end{equation}
and using the Clifford algebra, we obtain
\begin{equation}
    \left[
    g^{\mu\nu}\partial_{\mu}I\partial_{\nu}I+\mu^{2}
    \right]a(x)=0.
\end{equation}
For a nontrivial spinor amplitude, the Hamilton--Jacobi equation is therefore
\begin{equation}
    g^{\mu\nu}\partial_{\mu}I\partial_{\nu}I+\mu^{2}=0.
\end{equation}

Separate the action as
\begin{equation}
    I
    =
    -Et+W(r)+\Theta(\theta,\phi).
\end{equation}
The angular part is represented by the separation constant \(L^{2}\), namely
\begin{equation}
    \left(\partial_{\theta}\Theta\right)^{2}
    +
    \frac{1}{\sin^{2}\theta}
    \left(\partial_{\phi}\Theta\right)^{2}
    =
    L^{2}.
\end{equation}
Using the inverse metric components, the Hamilton--Jacobi equation becomes
\begin{equation}
    -\frac{E^{2}}{g(r)}
    +
    g(r)\left(\frac{\dd W}{\dd r}\right)^{2}
    +
    \frac{L^{2}}{r^{2}}
    +
    \mu^{2}
    =
    0.
\end{equation}
Solving for the radial momentum gives
\begin{equation}
    k_{r}^{2}(r,E,L)
    =
    \left(\frac{\dd W}{\dd r}\right)^{2}
    =
    \frac{E^{2}}{g^{2}(r)}
    -
    \frac{1}{g(r)}
    \left(
    \mu^{2}+\frac{L^{2}}{r^{2}}
    \right).
\end{equation}
In the semiclassical angular-momentum approximation,
\begin{equation}
    L^{2}\simeq l(l+1).
\end{equation}
Thus
\begin{equation}
    k_{r}^{2}(r,E,l)
    =
    \frac{E^{2}}{g^{2}(r)}
    -
    \frac{1}{g(r)}
    \left[
    \mu^{2}+\frac{l(l+1)}{r^{2}}
    \right].
\end{equation}

\section{Dirac mode counting in the thin-film region}

The radial WKB quantization condition is
\begin{equation}
    n_{r}\pi
    =
    \int k_{r}(r,E,l)\,\dd r.
\end{equation}
The total number of modes below energy \(E\) is therefore
\begin{equation}
    n(E)
    =
    \frac{g_{\rm D}}{\pi}
    \int_{\rhp+\epsilon}^{\rhp+\epsilon+\delta}\dd r
    \int_{0}^{l_{\max}}
    (2l+1)
    k_{r}(r,E,l)\,\dd l,
\end{equation}
where \(g_{\rm D}\) is the Dirac degeneracy. For a four-component Dirac spinor one may take
\begin{equation}
    g_{\rm D}=4.
\end{equation}
For a single two-spin-state fermion sector one may instead set
\begin{equation}
    g_{\rm D}=2.
\end{equation}

The condition \(k_{r}^{2}\geq 0\) gives
\begin{equation}
    \frac{E^{2}}{g^{2}(r)}
    -
    \frac{1}{g(r)}
    \left[
    \mu^{2}+\frac{l(l+1)}{r^{2}}
    \right]
    \geq 0.
\end{equation}
Multiplying by \(g^{2}(r)\), we get
\begin{equation}
    E^{2}
    -
    g(r)\mu^{2}
    -
    g(r)\frac{l(l+1)}{r^{2}}
    \geq 0.
\end{equation}
Therefore
\begin{equation}
    l_{\max}(l_{\max}+1)
    =
    r^{2}
    \left[
    \frac{E^{2}}{g(r)}-\mu^{2}
    \right].
\end{equation}

Now define
\begin{equation}
    y=l(l+1).
\end{equation}
Then
\begin{equation}
    \dd y=(2l+1)\dd l.
\end{equation}
The angular integral becomes
\begin{equation}
    \int_{0}^{l_{\max}}
    (2l+1)
    k_{r}(r,E,l)\,\dd l
    =
    \int_{0}^{y_{\max}}
    \left[
    \frac{E^{2}}{g^{2}(r)}
    -
    \frac{\mu^{2}}{g(r)}
    -
    \frac{y}{g(r)r^{2}}
    \right]^{1/2}
    \dd y.
\end{equation}
Let
\begin{equation}
    A(r,E)
    =
    \frac{E^{2}}{g^{2}(r)}
    -
    \frac{\mu^{2}}{g(r)}.
\end{equation}
Let
\begin{equation}
    B(r)
    =
    \frac{1}{g(r)r^{2}}.
\end{equation}
Then
\begin{equation}
    y_{\max}
    =
    \frac{A(r,E)}{B(r)}.
\end{equation}
The integral is
\begin{equation}
    \int_{0}^{A/B}
    \left[A-By\right]^{1/2}\dd y
    =
    \frac{2}{3B}A^{3/2}.
\end{equation}
Substituting \(A\) and \(B\), we find
\begin{equation}
    \int_{0}^{l_{\max}}
    (2l+1)
    k_{r}(r,E,l)\,\dd l
    =
    \frac{2}{3}
    \frac{r^{2}}{g^{2}(r)}
    \left[
    E^{2}-\mu^{2}g(r)
    \right]^{3/2}.
\end{equation}
Therefore the number of Dirac modes is
\begin{equation}
    n(E)
    =
    \frac{2g_{\rm D}}{3\pi}
    \int_{\rhp+\epsilon}^{\rhp+\epsilon+\delta}
    \frac{r^{2}}{g^{2}(r)}
    \left[
    E^{2}-\mu^{2}g(r)
    \right]^{3/2}
    \dd r.
\end{equation}

Near the horizon,
\begin{equation}
    g(r)\rightarrow 0.
\end{equation}
Hence the mass term is subleading in the leading ultraviolet entropy. Thus
\begin{equation}
    \left[
    E^{2}-\mu^{2}g(r)
    \right]^{3/2}
    =
    E^{3}+O(g).
\end{equation}
The leading mode number is
\begin{equation}
    n(E)
    =
    \frac{2g_{\rm D}E^{3}}{3\pi}
    I_{Q}
    +O(g),
\end{equation}
where
\begin{equation}
    I_{Q}
    =
    \int_{\rhp+\epsilon}^{\rhp+\epsilon+\delta}
    \frac{r^{2}}{g^{2}(r)}\,\dd r.
\end{equation}

\section{Fermionic free energy and entropy at the RVB temperature}

For fermions, the free energy is
\begin{equation}
    F_{\rm D}
    =
    -\frac{1}{\beta}
    \sum_{s}\ln\left(1+e^{-\beta E_{s}}\right).
\end{equation}
In the continuum approximation, integration by parts gives
\begin{equation}
    F_{\rm D}
    =
    -\int_{0}^{\infty}
    \frac{n(E)}{e^{\beta E}+1}\,\dd E.
\end{equation}
For the RVB-corrected thermal atmosphere, we take
\begin{equation}
    \beta=\beta_{\rm RVB}=\frac{1}{\TRVB}.
\end{equation}
Substituting the leading mode number,
\begin{equation}
    F_{\rm D}^{\rm RVB}
    =
    -
    \frac{2g_{\rm D}I_{Q}}{3\pi}
    \int_{0}^{\infty}
    \frac{E^{3}}{e^{\beta_{\rm RVB}E}+1}\,\dd E.
\end{equation}
The standard fermionic integral is
\begin{equation}
    \int_{0}^{\infty}
    \frac{E^{3}}{e^{\beta E}+1}\,\dd E
    =
    \frac{7\pi^{4}}{120\beta^{4}}.
\end{equation}
Therefore
\begin{equation}
    F_{\rm D}^{\rm RVB}
    =
    -
    \frac{7g_{\rm D}\pi^{3}}{180}
    \frac{I_{Q}}{\beta_{\rm RVB}^{4}}.
\end{equation}
The entropy is
\begin{equation}
    S_{\rm D}^{\rm RVB}
    =
    \beta_{\rm RVB}^{2}
    \frac{\partial F_{\rm D}^{\rm RVB}}{\partial \beta_{\rm RVB}}.
\end{equation}
Using
\begin{equation}
    \frac{\partial}{\partial\beta_{\rm RVB}}
    \left(
    -\frac{1}{\beta_{\rm RVB}^{4}}
    \right)
    =
    \frac{4}{\beta_{\rm RVB}^{5}},
\end{equation}
we get
\begin{equation}
    S_{\rm D}^{\rm RVB}
    =
    \frac{7g_{\rm D}\pi^{3}}{45}
    \frac{I_{Q}}{\beta_{\rm RVB}^{3}}.
\end{equation}
Since \(\beta_{\rm RVB}^{-1}=\TRVB\), the Dirac-field entropy becomes
\begin{equation}
    S_{\rm D}^{\rm RVB}
    =
    \frac{7g_{\rm D}\pi^{3}}{45}
    I_{Q}
    \left[
    \frac{g'(\rhp)}{4\pi}+\Cres
    \right]^{3}.
\end{equation}

This is the central Dirac-field result before rewriting the cutoff in proper-distance form. The residue correction enters through the cubic factor of the temperature because the free energy of a massless fermionic atmosphere scales as \(T^{4}\), while entropy scales as \(T^{3}\).

\section{Near-horizon expansion and area form}

Assume that the horizon is nonextremal, so that
\begin{equation}
    g'(\rhp)\neq 0.
\end{equation}
Near the horizon,
\begin{equation}
    g(r)
    =
    g'(\rhp)(r-\rhp)
    +O\left((r-\rhp)^{2}\right).
\end{equation}
Therefore
\begin{equation}
    g^{2}(r)
    =
    \left[g'(\rhp)\right]^{2}(r-\rhp)^{2}
    +O\left((r-\rhp)^{3}\right).
\end{equation}
In the thin-film region,
\begin{equation}
    r^{2}
    =
    \rhp^{2}+O(r-\rhp).
\end{equation}
Thus
\begin{equation}
    I_{Q}
    =
    \int_{\rhp+\epsilon}^{\rhp+\epsilon+\delta}
    \frac{r^{2}}{g^{2}(r)}\,\dd r
    =
    \frac{\rhp^{2}}{\left[g'(\rhp)\right]^{2}}
    \int_{\epsilon}^{\epsilon+\delta}
    \frac{\dd x}{x^{2}}
    +\cdots,
\end{equation}
where
\begin{equation}
    x=r-\rhp.
\end{equation}
The elementary integral is
\begin{equation}
    \int_{\epsilon}^{\epsilon+\delta}
    \frac{\dd x}{x^{2}}
    =
    \frac{1}{\epsilon}
    -
    \frac{1}{\epsilon+\delta}.
\end{equation}
Therefore
\begin{equation}
    I_{Q}
    =
    \frac{\rhp^{2}}{\left[g'(\rhp)\right]^{2}}
    \left[
    \frac{1}{\epsilon}
    -
    \frac{1}{\epsilon+\delta}
    \right]
    +\cdots.
\end{equation}
Substituting this into the entropy formula gives the coordinate-cutoff expression
\begin{equation}
    S_{\rm D}^{\rm RVB}
    =
    \frac{7g_{\rm D}\pi^{3}}{45}
    \frac{\rhp^{2}}{\left[g'(\rhp)\right]^{2}}
    \left[
    \frac{1}{\epsilon}
    -
    \frac{1}{\epsilon+\delta}
    \right]
    \left[
    \frac{g'(\rhp)}{4\pi}+\Cres
    \right]^{3}
    +\cdots.
\end{equation}

Now introduce the proper near-horizon cutoffs
\begin{equation}
    h
    =
    \int_{\rhp}^{\rhp+\epsilon}
    \frac{\dd r}{\sqrt{g(r)}}
    =
    2\sqrt{\frac{\epsilon}{g'(\rhp)}}+\cdots,
\end{equation}
and
\begin{equation}
    H
    =
    \int_{\rhp}^{\rhp+\epsilon+\delta}
    \frac{\dd r}{\sqrt{g(r)}}
    =
    2\sqrt{\frac{\epsilon+\delta}{g'(\rhp)}}+\cdots.
\end{equation}
These relations imply
\begin{equation}
    \frac{1}{\epsilon}
    =
    \frac{4}{g'(\rhp)h^{2}},
\end{equation}
and
\begin{equation}
    \frac{1}{\epsilon+\delta}
    =
    \frac{4}{g'(\rhp)H^{2}}.
\end{equation}
Hence
\begin{equation}
    \frac{1}{\epsilon}
    -
    \frac{1}{\epsilon+\delta}
    =
    \frac{4}{g'(\rhp)}
    \left(
    \frac{1}{h^{2}}-\frac{1}{H^{2}}
    \right).
\end{equation}
Define
\begin{equation}
    \Delta_{h}
    =
    \frac{1}{h^{2}}-\frac{1}{H^{2}}.
\end{equation}
Then
\begin{equation}
    I_{Q}
    =
    \frac{4\rhp^{2}}{\left[g'(\rhp)\right]^{3}}
    \Delta_{h}
    +\cdots.
\end{equation}
The proper-cutoff entropy becomes
\begin{equation}
    S_{\rm D}^{\rm RVB}
    =
    \frac{28g_{\rm D}\pi^{3}}{45}
    \frac{\rhp^{2}}{\left[g'(\rhp)\right]^{3}}
    \Delta_{h}
    \left[
    \frac{g'(\rhp)}{4\pi}+\Cres
    \right]^{3}.
\end{equation}

Using
\begin{equation}
    \Area=4\pi\rhp^{2},
\end{equation}
and factoring out the ordinary temperature \(g'(\rhp)/(4\pi)\), we get
\begin{equation}
    S_{\rm D}^{\rm RVB}
    =
    \frac{7g_{\rm D}}{2880\pi}
    \Area\,\Delta_{h}
    \left[
    1+\frac{4\pi\Cres}{g'(\rhp)}
    \right]^{3}.
\end{equation}
Since
\begin{equation}
    g'(\rhp)=\frac{\XiQ(\rhp)}{\rhp},
\end{equation}
the \(f(Q)\)-adapted expression is
\begin{equation}
    S_{\rm D}^{\rm RVB}
    =
    \frac{7g_{\rm D}}{2880\pi}
    \Area\,\Delta_{h}
    \left[
    1+\frac{4\pi\Cres\rhp}{\XiQ(\rhp)}
    \right]^{3}.
\end{equation}

If the outer boundary of the thin film is much farther than the brick-wall cutoff, then \(H\gg h\), and
\begin{equation}
    \Delta_{h}\simeq \frac{1}{h^{2}}.
\end{equation}
The entropy reduces to
\begin{equation}
    S_{\rm D}^{\rm RVB}
    \simeq
    \frac{7g_{\rm D}}{2880\pi}
    \frac{\Area}{h^{2}}
    \left[
    1+\frac{4\pi\Cres\rhp}{\XiQ(\rhp)}
    \right]^{3}.
\end{equation}
For a four-component Dirac field, \(g_{\rm D}=4\), so
\begin{equation}
    S_{\rm D}^{\rm RVB}
    \simeq
    \frac{7}{720\pi}
    \frac{\Area}{h^{2}}
    \left[
    1+\frac{4\pi\Cres\rhp}{\XiQ(\rhp)}
    \right]^{3}.
\end{equation}

For small \(\Cres\), the first-order expansion is
\begin{equation}
    S_{\rm D}^{\rm RVB}
    =
    \frac{7g_{\rm D}}{2880\pi}
    \Area\,\Delta_{h}
    \left[
    1+
    \frac{12\pi\Cres\rhp}{\XiQ(\rhp)}
    \right]
    +O(\Cres^{2}).
\end{equation}

\section{Quadratic model \(f(Q)=Q+\alpha Q^{2}\)}

Now consider the quadratic deformation
\begin{equation}
    f(Q)=Q+\alpha Q^{2}.
\end{equation}
Following the simple \(f(Q)\)-deformed black hole ansatz,
\begin{equation}
    g(r)=1-\frac{2M}{r}+\alpha r^{2}.
\end{equation}
Thus
\begin{equation}
    \psi_{Q}(r)=\alpha r^{2}.
\end{equation}
The function \(\XiQ(r)\) becomes
\begin{equation}
    \XiQ(r)
    =
    1+\psi_{Q}(r)+r\psi_{Q}'(r).
\end{equation}
Since
\begin{equation}
    \psi_{Q}'(r)=2\alpha r,
\end{equation}
we obtain
\begin{equation}
    \XiQ(r)=1+3\alpha r^{2}.
\end{equation}
The horizon mass relation is
\begin{equation}
    M(\rhp)
    =
    \frac{\rhp}{2}
    \left(
    1+\alpha\rhp^{2}
    \right).
\end{equation}
The derivative of the metric at the horizon is
\begin{equation}
    g'(\rhp)
    =
    \frac{1+3\alpha\rhp^{2}}{\rhp}.
\end{equation}
The RVB-corrected temperature is
\begin{equation}
    \TRVB(\rhp)
    =
    \frac{1+3\alpha\rhp^{2}}{4\pi\rhp}
    +
    \Cres.
\end{equation}

The geometric entropy follows from
\begin{equation}
    \frac{\dd S_{\rm geo}}{\dd \rhp}
    =
    \frac{2\pi\rhp(1+3\alpha\rhp^{2})}
    {1+3\alpha\rhp^{2}+4\pi\Cres\rhp}.
\end{equation}
For small \(\Cres\), this becomes
\begin{equation}
    \frac{\dd S_{\rm geo}}{\dd \rhp}
    =
    2\pi\rhp
    -
    8\pi^{2}\Cres
    \frac{\rhp^{2}}{1+3\alpha\rhp^{2}}
    +
    O(\Cres^{2}).
\end{equation}
Integrating term by term gives
\begin{equation}
    S_{\rm geo}^{(\alpha)}(\rhp)
    =
    \pi\rhp^{2}
    -
    8\pi^{2}\Cres
    \int^{\rhp}
    \frac{u^{2}}{1+3\alpha u^{2}}\,\dd u
    +
    O(\Cres^{2})
    +
    S_{0}.
\end{equation}
The integral is evaluated by writing
\begin{equation}
    \frac{u^{2}}{1+3\alpha u^{2}}
    =
    \frac{1}{3\alpha}
    \left[
    1-\frac{1}{1+3\alpha u^{2}}
    \right].
\end{equation}
Therefore
\begin{equation}
    \int
    \frac{u^{2}}{1+3\alpha u^{2}}\,\dd u
    =
    \frac{u}{3\alpha}
    -
    \frac{1}{3\alpha\sqrt{3\alpha}}
    \arctan\left(\sqrt{3\alpha}\,u\right).
\end{equation}
Choosing \(S_{0}=0\), we obtain
\begin{equation}
    S_{\rm geo}^{(\alpha)}(\rhp)
    =
    \pi\rhp^{2}
    -
    \frac{8\pi^{2}\Cres}{3\alpha}\rhp
    +
    \frac{8\pi^{2}\Cres}{3\alpha\sqrt{3\alpha}}
    \arctan\left(\sqrt{3\alpha}\,\rhp\right)
    +
    O(\Cres^{2}).
\end{equation}

The Dirac-field entropy in the same quadratic model is obtained by substituting
\begin{equation}
    \XiQ(\rhp)=1+3\alpha\rhp^{2}
\end{equation}
into the general formula. Thus
\begin{equation}
    S_{\rm D,\alpha}^{\rm RVB}
    =
    \frac{7g_{\rm D}}{2880\pi}
    \Area\,\Delta_{h}
    \left[
    1+
    \frac{4\pi\Cres\rhp}
    {1+3\alpha\rhp^{2}}
    \right]^{3}.
\end{equation}
For \(g_{\rm D}=4\), this becomes
\begin{equation}
    S_{\rm D,\alpha}^{\rm RVB}
    =
    \frac{7}{720\pi}
    \Area\,\Delta_{h}
    \left[
    1+
    \frac{4\pi\Cres\rhp}
    {1+3\alpha\rhp^{2}}
    \right]^{3}.
\end{equation}
For \(H\gg h\), we have
\begin{equation}
    S_{\rm D,\alpha}^{\rm RVB}
    \simeq
    \frac{7}{720\pi}
    \frac{\Area}{h^{2}}
    \left[
    1+
    \frac{4\pi\Cres\rhp}
    {1+3\alpha\rhp^{2}}
    \right]^{3}.
\end{equation}

The first-order residue expansion is
\begin{equation}
    S_{\rm D,\alpha}^{\rm RVB}
    =
    \frac{7g_{\rm D}}{2880\pi}
    \Area\,\Delta_{h}
    \left[
    1+
    \frac{12\pi\Cres\rhp}
    {1+3\alpha\rhp^{2}}
    \right]
    +
    O(\Cres^{2}).
\end{equation}
If \(\alpha\rhp^{2}\ll 1\), then
\begin{equation}
    \frac{1}{1+3\alpha\rhp^{2}}
    =
    1-3\alpha\rhp^{2}
    +
    O(\alpha^{2}).
\end{equation}
Thus
\begin{equation}
    S_{\rm D,\alpha}^{\rm RVB}
    =
    \frac{7g_{\rm D}}{2880\pi}
    \Area\,\Delta_{h}
    \left[
    1+
    12\pi\Cres\rhp
    -
    36\pi\alpha\Cres\rhp^{3}
    \right]
    +
    O(\Cres^{2},\alpha^{2}).
\end{equation}

\section{Total entropy branch}

In the test-field approximation, the total entropy can be written as the sum of the geometric entropy and the Dirac thermal-atmosphere entropy:
\begin{equation}
    S_{\rm total}^{\rm RVB}
    =
    S_{\rm geo}^{\rm RVB}
    +
    S_{\rm D}^{\rm RVB}.
\end{equation}
For the quadratic model, this gives
\begin{equation}
    S_{\rm total,\alpha}^{\rm RVB}
    =
    S_{\rm geo}^{(\alpha)}
    +
    S_{\rm D,\alpha}^{\rm RVB}.
\end{equation}
Substituting the explicit first-order expressions, we obtain
\begin{equation}
    \begin{aligned}
    S_{\rm total,\alpha}^{\rm RVB}
    =
    &
    \pi\rhp^{2}
    -
    \frac{8\pi^{2}\Cres}{3\alpha}\rhp
    +
    \frac{8\pi^{2}\Cres}{3\alpha\sqrt{3\alpha}}
    \arctan\left(\sqrt{3\alpha}\,\rhp\right)
    \\
    &
    +
    \frac{7g_{\rm D}}{2880\pi}
    \Area\,\Delta_{h}
    \left[
    1+
    \frac{12\pi\Cres\rhp}
    {1+3\alpha\rhp^{2}}
    \right]
    +
    O(\Cres^{2}).
    \end{aligned}
\end{equation}

The first line is the residue-corrected geometric entropy reconstructed from the first law. The second line is the regulated Dirac-field entropy. The latter is cutoff dependent and should be interpreted as matter-field entanglement or thermal-atmosphere entropy. As usual in brick-wall calculations, this divergent contribution can be absorbed into a renormalization of the gravitational coupling or fixed by a physical cutoff.

\section{Consistency checks}

First, when the residue correction vanishes,
\begin{equation}
    \Cres\rightarrow 0,
\end{equation}
the geometric entropy becomes
\begin{equation}
    S_{\rm geo}\rightarrow \frac{\Area}{4}.
\end{equation}
The Dirac-field entropy becomes
\begin{equation}
    S_{\rm D}
    \rightarrow
    \frac{7g_{\rm D}}{2880\pi}
    \Area\,\Delta_{h}.
\end{equation}
For \(g_{\rm D}=4\) and \(H\gg h\), this is
\begin{equation}
    S_{\rm D}
    \rightarrow
    \frac{7}{720\pi}
    \frac{\Area}{h^{2}}.
\end{equation}
This is the expected fermionic area-proportional brick-wall result.

Second, in the Schwarzschild limit,
\begin{equation}
    \alpha\rightarrow 0,
\end{equation}
we have
\begin{equation}
    \XiQ(\rhp)\rightarrow 1.
\end{equation}
Hence
\begin{equation}
    S_{\rm D}^{\rm RVB}
    \rightarrow
    \frac{7g_{\rm D}}{2880\pi}
    \Area\,\Delta_{h}
    \left(
    1+4\pi\Cres\rhp
    \right)^{3}.
\end{equation}
For small \(\Cres\), this becomes
\begin{equation}
    S_{\rm D}^{\rm RVB}
    =
    \frac{7g_{\rm D}}{2880\pi}
    \Area\,\Delta_{h}
    \left(
    1+12\pi\Cres\rhp
    \right)
    +
    O(\Cres^{2}).
\end{equation}

Third, the residue correction modifies the Dirac entropy through the ratio
\begin{equation}
    \frac{\TRVB}{\Tzero}
    =
    1+\frac{4\pi\Cres}{g'(\rhp)}
    =
    1+\frac{4\pi\Cres\rhp}{\XiQ(\rhp)}.
\end{equation}
Therefore
\begin{equation}
    S_{\rm D}^{\rm RVB}
    =
    S_{\rm D}^{(0)}
    \left(
    \frac{\TRVB}{\Tzero}
    \right)^{3}.
\end{equation}
This compact formula explains why the RVB residue enters the Dirac entropy cubically.

\section{Conclusion}

We have derived the Dirac-field entropy of a static spherically symmetric black hole in \(f(Q)\) gravity using the RVB residue-corrected temperature and the thin-film state-counting method. The leading WKB equation of the Dirac field reduces to the Hamilton--Jacobi equation, so the near-horizon pole structure of the radial momentum controls the mode density. After introducing a proper brick-wall cutoff, the Dirac-field entropy is

\begin{equation}
    S_{\rm D}^{\rm RVB}
    =
    \frac{7g_{\rm D}}{2880\pi}
    \Area\,\Delta_{h}
    \left[
    1+\frac{4\pi\Cres\rhp}{\XiQ(\rhp)}
    \right]^{3}.
\end{equation}

For the quadratic model \(f(Q)=Q+\alpha Q^{2}\), this becomes

\begin{equation}
    S_{\rm D,\alpha}^{\rm RVB}
    =
    \frac{7g_{\rm D}}{2880\pi}
    \Area\,\Delta_{h}
    \left[
    1+
    \frac{4\pi\Cres\rhp}
    {1+3\alpha\rhp^{2}}
    \right]^{3}.
\end{equation}

Thus the Dirac-field entropy remains area proportional after regularization, while the RVB residue modifies the coefficient through a temperature-renormalization factor. This provides a direct bridge between the residue-based \(f(Q)\) black hole thermodynamics and the traditional Dirac thin-film entropy method.

\end{document}